\documentclass[12pt]{article}
\def\L{\bar{\lambda}}
\def\bea{\begin{eqnarray}}
\def\eea{\end{eqnarray}}
\def\nn{\nonumber}
\def\ba{\begin{array}}
\def\ea{\end{array}}
\def\p{{\mbox{\boldmath $p$}}}
\def\pp{\p_\perp}
\def\hatp{{\hat{\mbox{\boldmath $p$}}}}
\def\hatpp{\hatp_\perp}
\def\r{{\mbox{\boldmath $r$}}}
\def\ddz{\frac{\partial}{\partial z}}
\def\E{{\hat{\cal E}}}
\def\O{{\hat{\cal O}}}
\def\Nab{{\mbox{\boldmath $\nabla$}}}
\def\H{{\rm H}}
\def\vsig{{\mbox{\boldmath $\sigma$}}}
\def\bpsi{{\mbox{\boldmath $\Psi$}}}
\def\ve{\varepsilon}
\begin{document}
\begin{center}
{\Large\bf Foldy-Wouthuysen transformation and} \\
{\Large\bf a quasiparaxial approximation scheme for} \\
{\Large\bf the scalar wave theory of light beams}

\bigskip

{\large Sameen Ahmed Khan,$^a$\footnote{{\em E-mail address}: khan@ce.fis.unam.mx}
Ramaswamy Jagannathan,$^b$\footnote{{\em E-mail address}: jagan@imsc.ernet.in} \\
Rajiah Simon$^b$\footnote{{\em E-mail address}: simon@imsc.ernet.in}}

\smallskip

$^a${\em Centro de Ciencias F\'{i}sicas, Universidad Nacional Aut\'{o}noma de
M\'{e}xico, \\
Apartado Postal 48-3,
Cuernavaca 62251, Morelos, M\'{E}XICO} \\
http://www.pd.infn.it/$\sim$khan/ \\
$^b${\em The Institute of Mathematical Sciences, \\
Central Institutes of Technology Campus, Tharamani,
Chennai - 600113, INDIA} \\
http://www.imsc.ernet.in/$\sim$jagan/ \\
http://www.imsc.ernet.in/$\sim$simon/ \\
http://www.imsc.ernet.in/$\sim$optics/ \\
\end{center}

\bigskip

\hrule

\bigskip

\noindent {\bf Abstract} \\

The Foldy-Wouthuysen iterative diagonalization technique is applied to the
Helmholtz equation to obtain a Hamiltonian description of the propagation of a
monochromatic quasiparaxial light beam through a system in which the refractive
index $n(x,y,z)$ varies about a background value $n_0$ such that
$\left|n(x,y,z)-n_0\right| \ll n_0$.  This technique is presented as an alternative
to the conventional method of series expansion of the radical.  Besides
reproducing all the traditional quasiparaxial terms, this method leads to
additional terms in the optical Hamiltonian. \\

\noindent {\em PACS}\,: 42.25.-p, 42.25.Bs  \\

\noindent {\em Keywords}: Scalar wave optics, Helmholtz equation, Beam propagation,
Quasiparaxial beams, Hamiltonian description, Mathematical methods of optics,
Foldy-Wouthuysen transformation.  \\

\hrule

\newpage

\section{Introduction}

In the traditional scalar wave theory for treating monochromatic quasiparaxial light
beam propagating along the positive $z$-axis, the $z$-evolution of the optical
wave function $\psi(\r)$ is taken to obey the Schr\"{o}dinger-like equation
\bea
i\L\frac{\partial\psi(\r)}{\partial z} = \hat{H}\psi(\r)\,,
\label{Sle}
\eea
where $\L$ $=$ $\lambda/2\pi$ $=$ $c/\omega$. The optical Hamiltonian $\hat{H}$ is
formally given by the radical
\bea
\hat{H} = -\left({n^2(\r)-\hatpp^2}\right)^{1/2}\,,
\eea
where $\hatp$ $=$ $-i\L\Nab$.  Then, one expands the square root in a series, with
suitable ordering, or symmetrization, of the resulting polynomials in the components
of $\r_\perp$ and $\hatpp$, if necessary, to get a Hermitian $\hat{H}$
\cite{GM}-\cite{D}.  Note that $n(\r)$ and $\hatpp$ $=$ $-i\L\Nab_\perp$ do
not commute.  Here we shall use the Foldy-Wouthuysen (FW) iterative diagonalization
procedure, well known in the treatment of the Dirac theory of electron \cite{BD} to
develop an alternative to this technique of obtaining the optical Hamiltonian.

\section{The traditional paraxial formalism}

Let the system we are considering have its optic axis along the $z$-direction and
let the refractive index of the medium, $n(\r)$, vary by a small amount around a
background, or mean, value $n_0$.  That is, $n(\r)$ $=$ $n_0-\ve(\r)$,
$\left|\ve(\r)\right|$ $\ll$ $n_0$, and $n_0^2-n^2(\r)$ $\approx$ $2n_0\ve(\r)$
$\ll$ $n_0^2$.  For a monochromatic quasiparaxial beam, with leading $z$-dependence
$\psi(\r)$ $\sim$ $\exp{\left(in_0z/\L\right)}$, we have $\left|\pp\right|$ $\ll$
$n_0$.  This means that $p_z \approx n_0$, and that all rays propagate almost
parallel to the positive $z$-axis.  Thus, the expansion of $\hat{H}$ in the small
parameters $\left|\pp\right|/n_0$ and $\ve(\r)/n_0$ is basically an expansion in
$1/n_0$.  We have
\bea
\hat{H} & = & -\left({n^2(\r)-\hatpp^2}\right)^{1/2} \nn \\
& \approx & -\left\{n_0^2 - \left(\hatpp^2
            + 2n_0\ve\right)\right\}^{1/2} \nn \\
& = & - n_0 + \left(\frac{1}{2n_0}\hatpp^2 + \ve\right) \
            + \frac{1}{2n_0}\left(\frac{1}{2n_0}\hatpp^2 + \ve\right)^2  \nn \\
&   & + \frac{1}{2n_0^2}\left(\frac{1}{2n_0}\hatpp^2 + \ve\right)^3
      + \frac{5}{8n_0^3}\left(\frac{1}{2n_0}\hatpp^2 + \ve\right)^4
      + \cdots\,\cdot
\label{convH}
\eea
If the terms up to first order in $1/n_0$ only are retained in $\hat{H}$, dropping
the terms proportional to the second and higher powers of $1/n_0$, one gets the
paraxial theory applicable to an ideal first order, or linear, system without any
aberrations, or nonlinearities.  To treat a nonlinear or aberrating system, to a
given order, one has to keep in $\hat{H}$ the terms proportional to powers of
$1/n_0$ up to the desired order of approximation.  To treat a given system the
corresponding Hamiltonian is chosen, up to desired order of accuracy, and the
integration of the optical Schr\"{o}dinger equation (\ref{Sle}) leads to the
required transfer (or Green's) function, for the evolution of $\psi(\r)$ across
the system, from the input transverse plane to the output transverse plane.

To arrive at the above formalism one usually starts with the scalar optical wave
equation
\bea
\left(\Nab^2
- \frac{n^2(\r)}{c^2}\frac{\partial^2}{\partial t^2}\right)
\Psi(\r,t) = 0\,,
\label{owe}
\eea
and specializes to the monochromatic case $\Psi(\r,t)$ $=$
$\psi(\r)\exp{\left(-i\omega t\right)}$.  Then, $\psi(\r)$ satisfies the Helmholtz
equation
\bea
\left(\Nab^2 + \frac{n^2(\r)}{\L^2}\right)\psi(\r) = 0\,.
\label{He}
\eea
Now, the optical Schr\"{o}dinger equation (\ref{Sle}) follows from
rewriting (\ref{He}) as
\bea
-\left(\L\frac{\partial}{\partial z}\right)^2\psi(\r) =
\left(n^2(\r) - \hatpp^2\right)\psi(\r)\,,
\eea
and then choosing the `square root' as
\bea
\left(i\L\ddz\right)\psi(\r) =
-\left(n^2(\r) - \hatpp^2\right)^{1/2}\,\psi(\r)\,,
\label{oSe}
\eea
corresponding to the requirement that the propagation be entirely in the positive
$z$-direction; if the propagation is in the negative $z$-direction, with
$\psi(\r) \sim \exp{\left(-in_0z/\L\right)}$, the right hand side of (\ref{oSe})
will have the opposite sign. As seen from (\ref{convH}), to first order in
$1/n_0$, (\ref{oSe}) becomes
\bea
i\L\frac{\partial\psi(\r)}{\partial z} \approx
\left(- n_0 + \frac{1}{2n_0}\hatpp^2 + \ve\right)\psi(\r)\,,
\label{parax}
\eea
the paraxial, or the parabolic, approximation to the Helmholtz equation.

It should be noted that the passage from (\ref{He}) to (\ref{oSe}) reduces the
original boundary value problem to a first order initial value problem in $z$.  This
reduction is of great practical value since it leads to the powerful system or the
Fourier optic approach \cite{G}  However, the above reduction process itself can
never be claimed to be rigorous or exact.  Hence there is room for alternative
procedures for the reduction.  Of course, any such reduction scheme is bound to lack
in rigor to some extent, and the ultimate justification lies only in the results
the scheme leads to.  The purpose of this note is to explore one such possibility
based on the analogy between (\ref{owe}) and the Klein-Gordon equation for a
spin-$0$ particle. Before beginning this exploration, it is useful to recount
briefly some of the other attempts to go beyond the paraxial regime.

\section{Beyond the paraxial approximation}

There have been several notable attempts to go beyond the paraxial approximation and,
in that process, to obtain a precise knowledge of the  meaning and accuracy of
paraxial wave optics itself. We highlight here only  some of these, and the
reader may consult these works for further references.

A significant early attempt in this regard is due to Lax {\em et al.} \cite{lax}.
These authors pointed out that the process of neglecting
${\rm grad}\,{\rm div}\, \vec{E}$ and seeking a solution that is plane polarized in
the same sense everywhere is simply incompatible with the exact Maxwell
equations. They developed an expansion procedure in powers of the beam
parameter $(w_0/\ell)$, where $w_0$ is the waist size and $\ell = 2\pi
w_0^2/\lambda$ is the diffraction length or (twice the) Rayleigh range of the
beam under consideration. In addition to showing that the zero-order field
obeyed the Maxwell system of equations, they developed the equations obeyed by
the higher-order corrections. The first-order correction was shown to be
longitudinal.

Agarwal and Pattanayak \cite{agarwal1} studied the propagation of Gaussian beam
in a simple [linear, homogeneous, and isotropic] dielectric using the angular
spectrum representation for the electric field, and showed  that the exact
solution consisted of the paraxial result plus higher-order non-Gaussian
correction terms. They demonstrated, in particular, that the second-order
correction term satisfied an equation consistent with the work of Lax {\em et al.}
cited above.  In another paper Agarwal and Lax \cite{agrawal2} examined the role of
the boundary condition in respect of the corrections to the paraxial approximation
of Gaussian beams. This work resolved the controversy between the work of Couture
and Belanger \cite{couture} and that of Agarwal and Pattanayak \cite{agarwal1}, by
tracing  it to the fact that the two works had used qualitatively different boundary
conditions for the correction terms: while Agarwal and Pattanayak had made the more
natural demand that the field distribution in the waist plane be strictly Gaussian
in the exact treatment, as in the paraxial case, Couture and Belanger had demanded
the on-axis field to be the same in both treatments.

A major step in directly connecting solutions of the paraxial equation to those
of the exact treatment was taken by W\"{u}nsche \cite{wunsche}, who showed that
it is possible to construct a linear {\em transition operator}  which transforms
arbitrary solutions of the paraxial equation into exact (monochromatic) solutions
of the scalar wave equation (Helmholtz equation). Indeed, W\"{u}nsche constructed
two such operators $T_1,\;T_2$ correspondng to two different boundary conditions
and noted, moreover, that the transition operator method is equivalent to the
{\em complete integration of the system of coupled differential equations of Lax}
{\em et al.}, restricted to the scalar case.  Cao and Deng \cite{cao-deng}
derived a simpler transform operator under the condition that the evanescent
waves can be ignored, and used this transform to study the corrections to the
paraxial approximation of {\em arbitrary} freely propagating beam. They verified
the consistency of their conclusions with those of the perturbation approach of
Lax {\em et al.} Subsequently, Cao \cite{cao} applied the method of Lax el al to
nonparaxial light beam propagating in a {\em transversely nonuniform} refractive
index medium, computed the correction terms in detail, and specialized the results
to the case of Gaussian beam propagating in transversely quadratic refractive index
media.

The transition operator method has been further extended by Alonso {\em et al.}
\cite{alonso1}.  The uncertainty product  has played a fundamental role in the
analysis of paraxial (both coherent and partially coherent) beams. Alonso and
Forbes \cite{alonso2} have recently generalized the uncertainty product to the
case of nonparaxial fields.

There are two aspects to the issue of going beyond paraxial optics. The first
one is to do with the spatial dependence alone, and hence applies to `scalar'
optics. The second one is to do with the vectorial nature of the light field
and, more specifically, with the fact that Maxwell's is a {\em constrained}
system of equations. The restriction ${\rm div}\,\vec{E} = 0$ (in free space
and in homogeneous dielectrics) demands that the spatial dependence of the
field, even in a transverse plane like the waist plane, cannot be chosen
independent of polarization. [Thus, an input plane-polarized plane wave going
through a thin spherical lens results, in the exit plane of the lens, in  a wave
 which is spherical and hence necessarily has spatially-varying polarization so
that the Poynting vector at all points in the exit plane points to the focus of
the lens.]    Though the work of Lax {\em et al.} pointed to both these aspects,
the subsequent ones noted above largely concentrated on the first.

Examining the fundamental Poincar\'{e} symmetry of the Maxwell system in the
front-form of Dirac, Mukunda {\em et al.} \cite{mukunda1} developed a
formalism which converts  a solution of the scalar wave equation  into a
corresponding (approximate) solution of the Maxwell system, resulting in a
generalization  of Fourier optics for vector electromagnetic beams
\cite{mukunda2}.  This formalism leads to simple-looking electromagnetic
Gaussian beams \cite{simon1}, and predicts a cross-polarization term
\cite{simon2} which is consistent with experimental observations
\cite{experiment}.  Further analysis of electromagnetic Gaussian beams has been
presented by Sheppard and Saghafi \cite{sheppard}, and by Chen {\em et al.}
\cite{chen}.

We describe below some preliminary results of an ongoing research on the use of
the FW transformations to study nonparaxial beams and the passage through
optical systems which are not necessarily paraxial (Gaussian).  There are two
primary reasons for our believing that this approach may have advantage over
the ones noted above.  First, the FW technique iteratively takes the field to a
new representation where the forward propagating components get progressively
{\em decoupled} from the backward propagating components.  Secondly, the FW
method appears ideally suited for the Lie algebraic approach \cite{D}.  Finally,
the FW technique generalizes to the vector case with very little extra effort, as
will be shown in a subsequent report.

\section{The Foldy-Wouthuysen formalism}

In the traditional scheme the purpose of expanding the Hamiltonian $\hat{H}$ $=$
$-\left(n^2(\r) - \hatpp^2\right)^{1/2}$ in a series using $1/n_0$ as the
expansion parameter is to understand the propagation of the quasiparaxial beam in
terms of a series of approximations (paraxial + nonparaxial).  Let us recall that
in relativistic quantum mechanics too one has a similar problem of understanding
the relativistic wave equation as the nonrelativistic approximation plus the
relativistic correction terms in the quasirelativistic regime.  For the Dirac
equation (which is first order in time) this is done using the FW transformation
leading to an iterative diagonalization technique.  For the Klein-Gordon equation
(which is second order in time) this is done using the same FW technique after
linearizing it with respect to time, and thus bringing it to a Dirac-like form,
following the Feschbach-Villars method \cite{BD}.  Nonrelativistic Schr\"{o}dinger
equation and the Klein-Gordon equation applicable in the cases of ion and
electron optics, when the spin is disregarded, have also been treated in a
similar way using the FW technique \cite{KJ,JK}.  The analogy between the
optical wave equation (\ref{owe}) and the Klein-Gordon equation suggests
naturally a similar technique for treating the scalar wave theory of light
beams.  Though the suggestion to employ the FW technique in the case of the
Helmholtz equation exists in the literature as a remark \cite{FM} it has not so
far been exploited to analyze the quasiparaxial approximations for any specific
beam optical system.

Written as a first order system, the Helmholtz equation reads
\bea
 &  & i\L\ddz
\left[\ba{c}
\psi(\r) \\
i\L\ddz\psi(\r)
\ea\right]
= \left[\ba{ccc}
0 &  & 1 \\
n^2(\r)-\hatpp^2 & & 0
\ea\right]
\left[\ba{c}
\psi(\r) \\
i\L\ddz\psi(\r)
\ea\right]. \nn \\
 &  &
\label{foHe}
\eea
Now, let us define
\bea
\bpsi^{(1)} = \left[
\ba{c}
\bpsi^{(1)}_+ \\
\bpsi^{(1)}_-
\ea\right] = \frac{1}{2}
\left[
\ba{c}
\psi - i\frac{\L}{n_0}\ddz\psi \\
\psi + i\frac{\L}{n_0}\ddz\psi
\ea\right]\,.
\label{transf}
\eea
The effect of the transformation in (\ref{transf}) for a quasiparaxial beam
moving in the forward $z$-direction is to separate the component which is `fast'
in $z$ from the one which is `slow'\,: there exists one linear combination of
$\psi$ and ${\partial\psi}/{\partial z}$ which varies rapidly in $z$ and another
which varies slowly in $z$, and the above transformation picks precisely these
components.  In our case of forward propagation, since $\psi(\r)$ $\sim$
$\exp{\left(in_0z/\L\right)}$, we have $\bpsi^{(1)}_+$ $\approx$ $\psi(\r)$ and
$\bpsi^{(1)}_+$ $\gg$ $\bpsi^{(1)}_-$.  In other words,  $\bpsi^{(1)}_+$ and
$\bpsi^{(1)}_-$ are, respectively, the {\it large} and the {\it small}
components of $\bpsi^{(1)}$.  Let us now rewrite the Helmholtz
equation (\ref{He}), or (\ref{foHe}), as
\bea
i\L\frac{\partial\bpsi^{(1)}}{\partial z} = \hat{\H}^{(1)}\bpsi^{(1)}\,,
\label{Heq}
\eea
with
\bea
\hat{\H}^{(1)} & = & -n_0\sigma_z + \E^{(1)} + \O^{(1)}\,, \nn \\
\E^{(1)} & = & \left(\frac{1}{2n_0}\hatpp^2 + \ve\right)\sigma_z\,, \nn \\
\O^{(1)} & = & \left(\frac{1}{2n_0}\hatpp^2 + \ve\right)
               \left(i\sigma_y\right)\,,
\label{H}
\eea
where $\sigma_y$ and $\sigma_z$ are, respectively, the $y$ and $z$ components of
the triplet of Pauli matrices, $\vsig$, namely,
\bea
\sigma_x =
\left[
\ba{cc}
0 & 1 \\
1 & 0
\ea\right]\,, \
\sigma_y =
\left[
\ba{lr}
0 & -i \\
i & 0
\ea\right]\,, \
\sigma_z =
\left[
\ba{lr}
1 & 0 \\
0 & -1
\ea
\right].
\eea
This form of the Helmholtz equation is analogous to the Feschbach-Villars form of the
Klein-Gordon equation.  Note that equation (\ref{Heq}) thus derived is algebraically
very similar to the Dirac equation. Like in the Dirac equation, $\hat{\H}^{(1)}$
is the sum of a leading diagonal term $-n_0\sigma_z$ (analogous to $mc^2\beta$),
a diagonal `even' term ${\E}^{(1)}$ which does not couple the large and the
small components of $\bpsi^{(1)}$, and the off-diagonal `odd' term ${\O}^{(1)}$
which couples them.  The even term ${\E}^{(1)}$ commutes with $\sigma_z$ and the
odd term ${\O}^{(1)}$ anticommutes with $\sigma_z$\,:$\E^{(1)}\sigma_z$ $=$
$\sigma_z\E^{(1)}$ and $\O^{(1)}\sigma_z$ $=$ $-\sigma_z\O^{(1)}$.  This perfect
analogy between the Dirac equation and (\ref{Heq}) enables us to use the standard
FW transformation technique to analyze (\ref{Heq}) in terms of paraxial and higher
order expansions with $1/n_0$ as the expansion parameter.  This technique works as
follows.

In $\hat{\H}^{(1)}$ the off-diagonal odd term is small compared to the diagonal part
$-n_0\sigma_z + \E^{(1)}$ in which the leading term is of order $n_0$. A series of
successive transformations of $\bpsi^{(1)}$, following a fixed recipe described below,
is applied to (\ref{Heq}) such that after each transformation the off-diagonal
odd term of the resulting equation becomes weaker, being of higher order in $1/n_0$.
If at any stage the odd term is considered weak enough to be neglected then one
can approximate the corresponding iterated Hamiltonian by keeping only its diagonal
part.  Thus, this iterative diagonalization scheme can be carried out systematically
up to any desired order in the expansion parameter $1/n_0$.  It is interesting to
note that (\ref{Heq}) corresponds already to the paraxial approximation (\ref{parax})
of the Helmholtz equation if the odd term $\O^{(1)}$ is dropped from
$\hat{\H}^{(1)}$, retaining only the diagonal part $-n_0\sigma_z + \E^{(1)}$.

The first FW transformation is
\bea
\bpsi^{(2)} & = & \exp\left(-\sigma_z\O^{(1)}/2n_0\right)\bpsi^{(1)} \nn \\
  & = & \exp\left[-\frac{1}{2n_0}
        \left(\frac{1}{2n_0}\hatpp^2 + \ve\right)
        \sigma_x\right]\bpsi^{(1)}\,.
\eea The result of this transformation is to turn (\ref{Heq}) into
\bea
i\L\frac{\partial\bpsi^{(2)}}{\partial z} = \hat{\H}^{(2)}\bpsi^{(2)}\,,
\label{order2}
\eea
with
\bea
\hat{\H}^{(2)} & = & -n_0\sigma_z + \E^{(2)} + \O^{(2)}\,, \nn \\
\E^{(2)} & = & \E^{(1)}
               - \frac{1}{2n_0}\sigma_z\left(\O^{(1)}\right)^2
               -\frac{1}{8n_0^2} \left[\O^{(1)}, \left[\O^{(1)},\E^{(1)}\right]
               + i\L\frac{\partial\O^{(1)}}{\partial z}\right] + \cdots\,, \nn \\
\O^{(2)} & = & -\frac{1}{2n_0}
               \sigma_z\left(i\L\frac{\partial\O^{(1)}}{\partial z}\right)
               -\frac{1}{2n_0}\sigma_z\left[\O^{(1)},\E^{(1)}\right]
               + \cdots . \nn \\
 &   &
\label{eoexp}
\eea
Note that in $\hat{\H}^{(1)}$ the odd term $\O^{(1)}$ is of order
$\left(1/n_0\right)$.  In $\hat{\H}^{(2)}$ the odd term $\O^{(2)}$ is of order
$\left(1/n_0\right)^2$.  Hence in (\ref{order2}) the odd part is weaker than the
diagonal part, by one more order, compared to (\ref{Heq}).  Further, note that
the basic algebraic structure, $\E^{(2)}\sigma_z$ $=$ $\sigma_z\E^{(2)}$ and
$\O^{(2)}\sigma_z$ $=$ $-\sigma_z\O^{(2)}$, is preserved by the iteration.

The second FW transformation is
\bea
\bpsi^{(3)} = \exp\left(-\sigma_z\O^{(2)}/2n_0\right)\bpsi^{(2)}\,.
\eea
The result of this transformation is to turn (\ref{order2}) into
\bea
i\L\frac{\partial\bpsi^{(3)}}{\partial z} = \hat{\H}^{(3)}\bpsi^{(3)}\,,
\eea
with
\bea
\hat{\H}^{(3)} = -n_0\sigma_z + \E^{(3)} + \O^{(3)}\,,
\eea
where $\E^{(3)}$ and $\O^{(3)}$ are obtained by the replacements
\bea
\E^{(3)} & = & \E^{(2)}\left(\E^{(1)} \rightarrow \E^{(2)}\,,\,
               \O^{(1)} \rightarrow \O^{(2)}\right)\,, \nn \\
\O^{(3)} & = & \O^{(2)}\left(\E^{(1)} \rightarrow \E^{(2)}\,,\,
               \O^{(1)} \rightarrow \O^{(2)}\right)\,.
\eea
This happens because the expressions in (\ref{eoexp}) follow as a result of the
general algebraic properties of the even and odd operators, independent of the
specific expressions for $\E^{(1)}$ and $\O^{(1)}$ given in (\ref{H}).  Further,
$\E^{(3)}\sigma_z$ $=$ $\sigma_z\E^{(3)}$ and $\O^{(3)}\sigma_z$ $=$
$-\sigma_z\O^{(3)}$, and $\O^{(3)}$ is of order $\left(1/n_0\right)^3$.  Thus the
odd part of $\hat{\H}^{(3)}$ is weaker than the odd part of $\hat{\H}^{(2)}$ by one
more order.  Now, it is straightforward to see how this sequence of FW
transformations can be continued up to any desired stage, making the odd part of
the resulting $\hat{\H}^{(\cdot)}$ weaker by one more order at each stage.

If the FW transform process is stopped at, say, the $j$-th stage then one would
have arrived at
\bea
i\L\frac{\partial\bpsi^{(j)}}{\partial z} = \hat{\H}^{(j)}\bpsi^{(j)}\,,
\eea
with
\bea
\bpsi^{(j)} = \left(\ba{c}
                  \bpsi^{(j)}_+ \\
                  \bpsi^{(j)}_-
                  \ea\right)\,, \quad
\hat{\H}^{(j)} = -n_0\sigma_z+\E^{(j)}+\O^{(j)}\,,
\eea
where $\O^{(j)}$ is of order $\left(1/n_0\right)^j$.  It is important to note that
each FW transformation preserves and improves the property that the upper component
of $\bpsi^{(\cdot)}$ is large compared to its lower component\,: $\bpsi^{(j)}_+$
$\gg$ $\bpsi^{(j)}_-$.  In view of this, we can drop the odd part $\O^{(j)}$ from
$\hat{\H}^{(j)}$, as negligible compared to the diagonal part $-n_0\sigma_z+\E^{(j)}$,
and write
\bea
i\L\frac{\partial\psi(\r)}{\partial z} \approx \hat{\cal H}\psi(\r)\,,
\qquad
\hat{\cal H} = -n_0+\E^{(j)}_{11}\,,
\label{orderj}
\eea
where $\E^{(j)}_{11}$ is the $11$ matrix element of $\E^{(j)}$ and $\bpsi^{(j)}_+$
has been simply relabeled $\psi$.  Equation (\ref{orderj}) is the $j$-th order
approximation to the Helmholtz equation in this approach.

As already noted the first order approximation corresponds to the usual
paraxial theory.  In this case, equation (\ref{orderj}) becomes
\bea
i\L\frac{\partial\psi(\r)}{\partial z} & \approx & \hat{\cal H}\psi(\r)\,,
\nn \\
\hat{\cal H} & = & -n_0+\E^{(1)}_{11}
               = -n_0+\frac{1}{2n_0}\hatpp^2+\ve\,.
\eea
Let us now look at the second order approximation.  From (\ref{H}),
(\ref{order2}), (\ref{eoexp}), and (\ref{orderj}), we have, keeping terms up to order
$\left(1/n_0\right)^5$,
\bea
i\L\frac{\partial\psi(\r)}{\partial z} & \approx & \hat{\cal H}\psi(\r)\,,
\nn \\
\hat{\cal H} & = & -n_0+\E^{(2)}_{11} \nn \\
   & = & - n_0 + \left(\frac{1}{2n_0}\hatpp^2 + \ve\right)
         + \frac{i\L}{16n_0^3}
         \left[\hatpp^2\,,\,\frac{\partial\ve}{\partial z}\right] \nn \\
   &   & + \frac{1}{2n_0}\left(\frac{1}{2n_0}\hatpp^2 + \ve\right)^2
         + \frac{1}{2n_0^2}\left(\frac{1}{2n_0}\hatpp^2 + \ve\right)^3
         + \ldots \nn \\
   & = & - n_0 + \left(\frac{1}{2n_0}\hatpp^2 + \ve\right) \nn \\
   &   &  + \frac{\L^2}{16n_0^3}\frac{\partial}{\partial z}
         \left(\hatpp\cdot\Nab_\perp\ve + \Nab_\perp\ve\cdot\hatpp\right)
         + \ldots\,\cdot
\label{calH}
\eea
Comparing (\ref{calH}) with the traditional expansion in (\ref{convH}) it is
clear that $\hat{\cal H}$ has the first few terms of the traditional Hamiltonian
$\hat{H}$ correctly, plus an extra term (the commutator term).  To get the higher
order terms of $\hat{H}$ in $\hat{\cal H}$ with the correct coefficients one will
have to consider approximations beyond the second order.  We assume, for
consistency, that the derivatives of $\ve(\r)$ are also small compared to $n_0$.

One may note that the FW scheme automatically leads to a Hermitian Hamiltonian
without need for any further ordering of its noncommuting components.

\section{Concluding remarks}

It is interesting that the extra commutator term
$\frac{i\L}{16n_0^3}\left[\hatpp^2\,,\,\frac{\partial\ve}{\partial z}\right]$
in (\ref{calH}) contributes a correction to the optical Hamiltonian, even at
the `paraxial level', when the refractive index of the medium suffers both
longitudinal and transverse inhomogeneities.  Such a $z$-derivative term is not
natural to the traditional power series expansion.  This commutator term is
what survives in the commutator term $-\frac{1}{8n_0^2} \left[\O^{(1)},
\left[\O^{(1)},\E^{(1)}\right] + i\L\frac{\partial\O^{(1)}}{\partial z}\right]$
in the expression for $\E^{(2)}$ in (\ref{eoexp}).  In the Foldy-Wouthuysen
formalism of the Dirac theory the corresponding commutator term is responsible
for the correct explanation of the spin-orbit energy (including the Thomas precession
effect) and the Darwin term (attributed to the zitterbewegung) (see Sectioin 4.3 of
\cite{BD}).  Similarly in the nonrelativistic reduction and interpretation of the
Klein-Gordon equation using the Foldy-Wouthysen transformation theory such a
commutator term corresponds to the Darwin term correcting the classical electrostatic
interation of a point charge in analogy to the zitterbewegung of the Dirac electron
(see Section 9.7 of \cite{BD}).  In the quantum theory of beam optics of charged
Klein-Gordon and Dirac particles \cite{KJ,JK} the corresponding terms add to the
Hamiltonian the lowest order quantum corrections to the classical aberration
terms.   In view of this analogy it should be of interest to study the effect of
this correction term to the optical Hamiltonian,
$\frac{i\L}{16n_0^3}\left[\hatpp^2\,,\,\frac{\partial\ve}{\partial z}\right]$,
on the propagation of Gaussian beams in a parabolic index medium whose focusing
strength is modulated in the axial variable $z$, however tiny it may be compared
to the classical terms.

Other questions naturally suggested by the above preliminary report are: the
issue of convergence in respect of the series expansion resulting from the FW
method and the boundary condition the series satisfies in relation to the
paraxial result in any special plane like the waist plane of the beam.  The
precise relation between the FW series and the results of the other approaches
recounted in Section 3 are not immediately clear at the moment, but it is an
important issue worth investigating.  We hope to return to these issues elsewhere.

The authors would like to thank the Referees for some insightful comments
leading to improved clarity of presentation.

\end{document}